\documentclass[]{jfm}

\usepackage{graphicx}
\usepackage{newtxtext}
\usepackage{newtxmath}
\usepackage{natbib}
\usepackage{hyperref}
\usepackage{float}
\usepackage{upgreek}

\hypersetup{
    colorlinks = true,
    urlcolor   = blue,
    citecolor  = blue,
}

\newcommand{\RomanNumeralCaps}[1]
\linenumbers

\title{Marangoni-driven freezing dynamics of supercooled binary droplets}

\author{Feng Wang\aff{1}, Hao Zeng\aff{1}, Yihong Du\aff{1}, Xinyu Tang\aff{1}, \and Chao Sun\aff{1,2}\corresp{\email{chaosun@tsinghua.edu.cn}}}

\affiliation{\aff{1}New Cornerstone Science Laboratory, Center for Combustion Energy, Key Laboratory for Thermal Science and Power Engineering of Ministry of Education,
Department of Energy and Power Engineering, Tsinghua University, 100084 Beijing, China
\aff{2}Department of Engineering Mechanics, School of Aerospace Engineering, Tsinghua University, Beijing 100084, China
}

\begin{document}
\maketitle

\begin{abstract}
Solidification of droplets is of great importance to various technological applications, drawing considerable attention from scientists aiming to unravel the fundamental physical mechanisms. In the case of multicomponent droplets undergoing solidification, the emergence of concentration gradients may trigger significant interfacial flows that dominate the freezing dynamics. Here, we experimentally investigate the fascinating interfacial freezing dynamics of supercooled ethanol-water droplets, accompanied with the migration and growth of massive ice particles. We reveal that these unique freezing dynamics are driven by solidification-induced solutal Marangoni flow within the droplets. Our model, which incorporates the temperature- and concentration-dependent properties of the ethanol-water mixture, quantitatively predicts both the migration velocity and the growth rate of the ice particles. The former is determined by the solutal Marangoni flow velocity, while the latter is governed by a balance between the latent heat release and the enhanced thermal dissipation by the Marangoni flow. Moreover, we show that the final wrapping state of droplets can be modulated by the concentration of ethanol. Our findings may pave the way for novel insights into the physicochemical hydrodynamics of multicomponent liquids undergoing phase transitions.
\end{abstract}


%{\bf MSC Codes }  {\it(Optional)} Please enter your MSC Codes here

\section{Introduction}
\label{sec:headings}

Solidification of droplets is ubiquitous in both natural environments and industrial applications. The accumulation of frozen droplets on artificial surfaces can lead to a range of detrimental consequences to aviation safety \citep{cebeci2003aircraft}, maritime operations \citep{dehghani2017sea} and wind turbines \citep{makkonen2001modelling} in the cold regions. Substantial efforts have been dedicated to enhancing our understanding of the intricate dynamics of freezing droplet \citep{kant2020fast,koldeweij2021initial,grivet2022contact,tembely2019comprehensive,zadravzil2006droplet,kavuri2023freezing,kavuri2025evaporation} covering a wide array of factors, such as the influence of environmental medium \citep{lyu2023liquid,lambley2023freezing,graeber2017spontaneous}, the inertia of droplets \citep{hu2020frozen,thievenaz2019solidification,thievenaz2020retraction}, the composition of droplets \citep{zeng2023evaporation,jiang2024three,kharal2023unidirectional}, and the characteristics of substrates \citep{lolla2022arrested,wang2023droplet,de2018self}.

\begin{figure*}
\centerline{\includegraphics[width=1.0\textwidth]{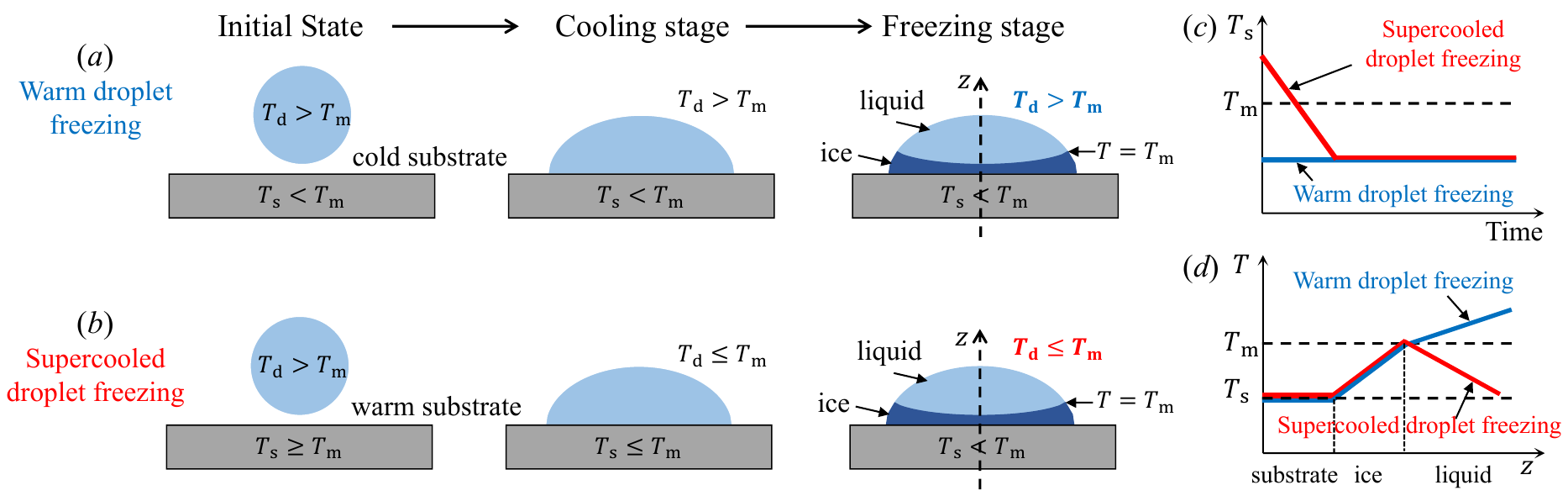}}%
\caption{\label{fig1} Schematic diagrams illustrating droplet freezing dynamics. \textbf{(a)} For warm droplets, the liquid phase temperature $T_\text{d}$ remains above the freezing point $T_\text{m}$, so the released latent heat at the freezing front is primarily dissipated via ice-phase conduction. \textbf{(b)} In contrast, supercooled droplets (with $T_\text{d}<T_\text{m}$) enable additional heat dissipation through the supercooled liquid during freezing. \textbf{(c)} Substrate temperature profiles over time comparing warm (blue) and supercooled (red) droplet freezing scenarios. \textbf{(d)} Temperature profiles along the dashed line in (a,b) during solidification for warm (blue) and supercooled (red) droplet freezing, showing distinct thermal behaviors.}
\end{figure*}

Based on the temperature of the remaining liquid during solidification, droplet freezing systems can be categorized into two types: (1) warm droplet freezing (Fig.~\ref{fig1}a) and (2) supercooled droplet freezing (Fig.~\ref{fig1}b). In the case of warm droplet freezing, the initial droplet temperature $T_\text{d}$ is higher than the freezing point $T_\text{m}$, while the initial substrate temperature $T_\text{s}$ is lower than the freezing point $T_\text{m}$, as illustrated by the blue solid line in Fig.~\ref{fig1}(c). Owing to the presence of potential ice nucleation sites on the undercooled substrate, such as pre-existing frost droplets and ice layers \citep{zeng2024extended,lolla2022arrested}, the droplet starts to solidify immediately upon contact with the undercooled substrate. During this process, the temperature of the remaining liquid stays above the freezing point \citep{wang2024self}, as demonstrated by the blue solid line in Fig.~\ref{fig1}(d). Consequently, the latent heat released at the ice front can only be consumed through heat conduction via the ice phase. In contrast, when it comes to the freezing of supercooled droplets, at the initial state, both the droplet temperature $T_\text{d}$ and the substrate temperature $T_\text{s}$ are higher than the freezing point $T_\text{m}$. Subsequently, the substrate is gradually cooled to a temperature below the freezing point, as indicated by the red solid line in Fig.~\ref{fig1}(c). Simultaneously, the droplet is also cooled and enters a supercooled state \citep{chu2024interfacial,schremb2017ice} due to the ice nucleation energy barrier. Then, the supercooled droplet undergoes a transition from its supercooled state to a mixture of ice and liquid water at the freezing temperature, which is referred to as the recalescence stage \citep{sebilleau2021air}. The red solid line in Fig.~\ref{fig1}(d) represents the typical temperature distribution along the centerline of the freezing supercooled droplet, as marked by the dashed line in Fig.~\ref{fig1}(b). Throughout solidification, the temperature of the remaining liquid remains supercooled \citep{schremb2017ice} As a result, the latent heat released at the ice front can be consumed not only by heat conduction through the ice phase but also by heat transfer within the supercooled liquid.

The freezing dynamics varies significantly between supercooled droplets and warm droplets on cold surfaces. For warm pure water droplets, a unidirectional ice front advancing away from the cold substrate was observed \citep{marin2014universality}. This is attributed to the dominant heat conduction through the ice phase, which dissipates the latent heat released during solidification. Conversely, during the solidification of supercooled water droplets on a cold wall, dendritic freezing dynamics were identified \citep{schremb2017ice}. This process requires consideration of heat conduction in both the supercooled liquid and the adjacent solid substrate. Similarly, the freezing dynamics of warm saline droplets resemble those of warm pure water droplets, with a slight difference in the tip angle formed at the top of the droplets \citep{boulogne2020drop}. This difference arises from the volume expansion during solidification \citep{marin2014universality}. However, for supercooled saline droplets, an unexpected interfacial ice sprouting dynamics has been reported \citep{chu2024interfacial}. This unique phenomenon is attributed to the formation of a brine film during the nucleation and recalescence process, which is accompanied by salt rejection.

To further explore the impact of composition changes on freezing dynamics, the role of the solutal Marangoni effect becomes particularly interesting. If the composition rejected near the ice front can significantly alter surface tension, a strong solutal Marangoni effect may arise within the droplets, potentially influencing their freezing dynamics. \citet{wang2024self} observed this effect in warm ethanol-water droplets on a cold substrate, reporting the self-lifting freezing dynamics. The solutal Marangoni flow, driven by ethanol enrichment at the ice front, created a pronounced dynamic growth angle at the liquid-solid-gas tri-junction, resulting in the nearly doubling  of the droplets' final height due to geometrical effects \citep{anderson1996case}.

Building on these understandings, we now turn to the question of what occurs during the solidification of supercooled ethanol-water droplets on a cold surface. The nucleation and growth of ice crystals accompanied with the rejection of ethanol may involve more intricate physical phenomena than those in previous work. The presence of Marangoni flow could significantly enhance convective heat transfer and mass transport, leading to novel freezing dynamics in supercooled ethanol-water droplets.

Here, we investigate experimentally the fascinating migration and growth dynamics of ice particles at the interface of supercooled ethanol-water droplets. This phenomenon is propelled by solutal Marangoni flow \citep{wang2024self}, which arises from the localized enrichment of ethanol near the ice front \citep{dedovets2018five}. Our findings reveal that these unique freezing dynamics are driven by solidification-induced solutal Marangoni flow. Using a model that accounts for the property variations of the ethanol-water mixture, we achieve quantitative agreement with experiments on two key aspects: the ice particle migration velocity, predicted by the solutal Marangoni flow speed, and its growth rate, modeled by balancing latent heat release against solutal-Marangoni-enhanced thermal dissipation. This novel interfacial freezing process culminates in the formation of beautiful patterns covered by ice particles \citep{ahmadi2019soap}, the distinctive dimensions of which are found to depend on the concentration of ethanol.

The paper is organized as follows. Section \ref{sec:2} describes the experimental set-up and the properties of the binary droplets. Section \ref{sec:3} describes the experimental results, providing a general picture of the interfacial freezing dynamics of supercooled binary droplets.  In Section \ref{sec:4}, we focus on the interfacial temperature evolution of the supercooled binary droplets. In Section \ref{sec:5}, we elucidate the Marangoni-driven dynamics of the ice particles at the droplet-air interface. Finally, we end with conclusions and discussions in Section \ref{sec:6}.

\section{Materials and methods}
\label{sec:2}
\begin{figure*}
\centerline{\includegraphics[width=0.8\textwidth]{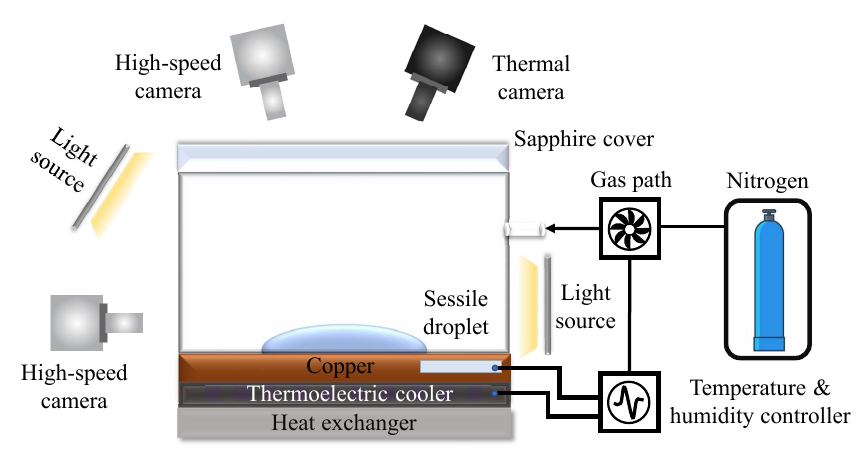}}%
\caption{\label{fig2} Schematic view of the experimental setup. In the experiments, the droplet is gently deposited onto the copper substrate, with both the droplet and substrate preconditioned to room temperature. Subsequently, the substrate undergoes progressive cooling down to $T_\text{\text{s}}=-20 ^\circ$C by a semiconductor cooler.}
\end{figure*}

Experiments are conducted in a custom-built transparent environmental chamber with internal dimensions of 8 cm $\times$ 8 cm $\times$ 8 cm. This chamber allows for humidity control and temperature monitoring, as depicted in Fig.~\ref{fig2}. The droplets used in the experiments are either deionized water or ethanol-water solutions. The ethanol, supplied by Shanghai Titan Scientific Co.Ltd (General-reagent, $\geq$ 99.7$\%$), is mixed with deionized water (18.2 $\rm{M\Omega\cdot{cm}}$ electrical conductivity) from a Milli-Q water purification system (Merck, Germany). Microliter droplets are generated using a microsyringe and dispensing needle. The substrate is gradually cooled by an embedded semiconductor cooler. A heat exchanger connected to a cold reservoir is attached to the opposite side of the semiconductor to maintain energy balance. A thermocouple, embedded 0.5 mm below the substrate's upper surface, measures its temperature. Connected to a PID controller, the thermocouple regulates the power supply to the semiconductor cooler, ensuring a variation of less than 0.2 K during experiments.

In the experiments, droplets with various volume ($V_0$) and volume fraction of ethanol ($c_0$) are gently deposited onto the nickel-plated copper substrate at room temperature ($T_0=25~^\circ$C). Then, the substrate is slowly cooled by an embedded peltier cooler down to $T_\text{s}=-20~^\circ$C, and the droplet is gradually cooled into a supercooled state before the suddenly nucleation of ice crystal \citep{chu2024interfacial}. The freezing dynamics of the droplets are recorded by high-speed cameras (Photron Nova S12) from both the top-down view and side view, and the temperature evolution is recorded by an IR camera (Telops FAST L200, Stirling-cooled MCT detector) from the top-down view.

The physical properties of the binary droplets employed in this work are summarized in Tab.~\ref{tab:my_label}. The freezing point of the ethanol-water mixture, along with its surface tension and density at $T=20~^\circ$C were determined by fitting polynomial functions to experimental data \citep{dean1999lange,kuchuk2012physicochemical,khattab2012density}. The viscosity of the ethanol-water mixture at $T=-20~^\circ$C was obtained by interpolating the measurement data from the literature \citep{halfpap1982viscosity}.

\begin{table}
    \centering
    \tabcolsep=0.5cm
    \begin{tabular}{ccccc}
   
       Concentration  & Freezing point & Surface tension & Density & Viscosity \\
        $c_{0}$ (v/v)& $T_{f}$ ($^{\circ}$C) & $\gamma$ (mN/m) & $\rho$ (g/cm$^3$) & $\mu$ (mPa$\cdot$s) \\ 
        0 &  0&   72.41 & 1.00 & 4.29 \\
        0.05 &  $-1.61$& 60.30 & 0.991 & 5.77  \\
        0.10 &  $-3.35$&  51.49 & 0.985 & 8.44 \\
        0.15 & $-5.38$&  45.14 & 0.979 & 11.27\\
        0.20 & $-7.79$&  40.79 & 0.974 & 14.40\\
        0.37 & $-20$&  34.20 & 0.953 & 25.67 \\
    \end{tabular}
    \caption{Properties of ethanol-water droplets: surface tension and density at $T=20~^\circ$C, and viscosity at $T=-20~^\circ$C.}
     \label{tab:my_label}
\end{table}

\section{Observation of the interfacial freezing dynamics} 
\label{sec:3}

\begin{figure*}
\centerline{\includegraphics[width=0.8\textwidth]{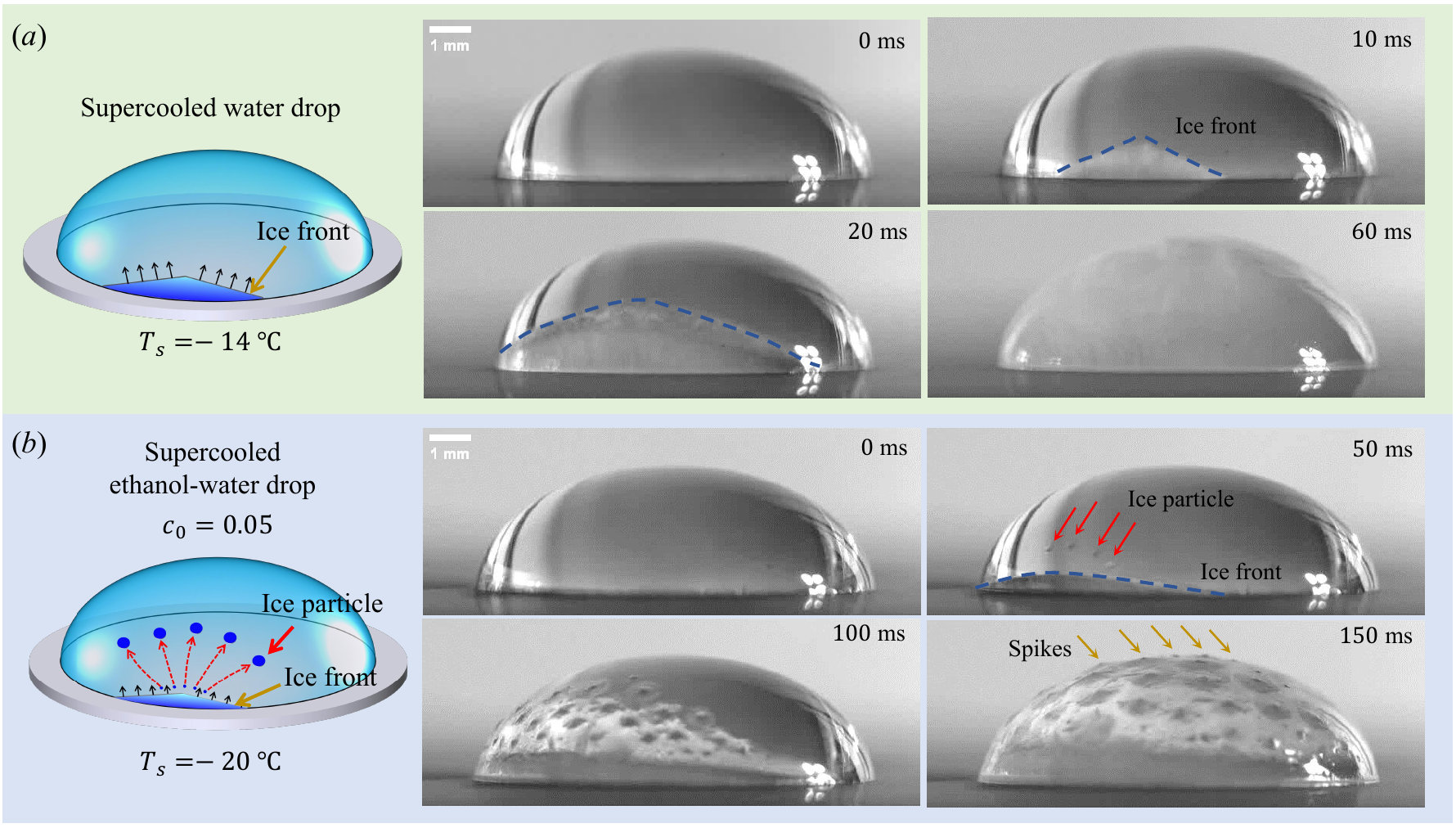}}%
\caption{\label{fig3} Side view for the freezing dynamics of supercooled droplets. \textbf{(a)} In the case of pure water droplets ($V_0=150~\upmu$L), the explosive nucleation phenomena (recalescence) is observed, when the substrate temperature $T_\text{\text{s}}$ is cooled down to about $-14 ^\circ$C. \textbf{(b)} By contrast, the ethanol-water droplets ($c_0=0.05$, $V_0=150~\upmu$L) can remain unfrozen for several minutes, with the substrate temperature maintaining $T_\text{\text{s}}=-20 ^\circ$C. Subsequently, dispersed ice particles are observed at the droplet-air interface, migrating uni-directionally and growing slowly, which is quite different from that of pure water case. }
\end{figure*}

Figure~\ref{fig3} presents the typical freezing dynamics of supercooled droplets (Movie S1). For a pure water droplet ($V_0=150~\upmu$L, Fig.~\ref{fig3}a), the rapid recalescence process is observed when the substrate temperature approaches the nucleation temperature $T_\text{s} \approx -14 ^\circ$C. The dendritic ice front, as indicated by the blue dashed line in Fig.~\ref{fig3}(a), advances rapidly with the velocity of approximately 10 cm/s, aligning with prior studies \citep{schremb2017ice}. In contrast, for an ethanol-water droplet ($V_0=150~\upmu$L, $c_0=0.05$, Fig.~\ref{fig3}b), the initial formation of dendritic ice front is also observed, as indicated by the blue dash line in Fig.~\ref{fig3}b. However, a surprising phenomenon occurs: ice particles detach and move away from the ice front at the droplet-air interface (indicated by red arrows), migrating toward the droplet's opposite side (as shown by the red dashed line in the schematic of Fig.~\ref{fig3}b). These ice particles travel faster than the advancing ice front, with their amount and size progressively increasing, forming spikes at the droplet-air interface (as indicated by the red arrows at $t=150$ ms). 

\begin{figure*}
\centerline{\includegraphics[width=1.0\textwidth]{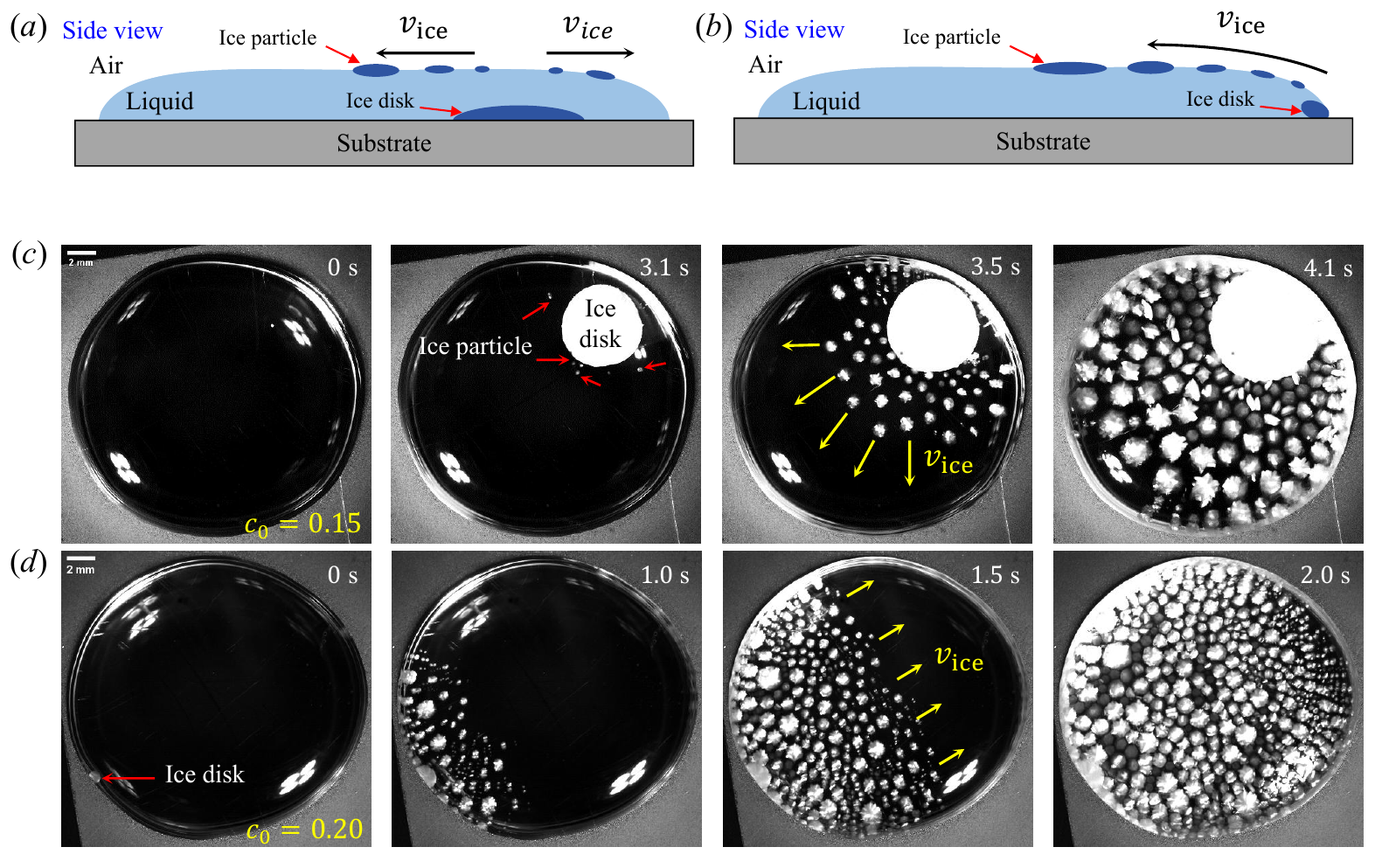}}%
\caption{\label{fig4} Top-down view for the freezing dynamics of supercooled ethanol-water droplets. \textbf{(a,b)} Sketch for two typical freezing processes observed in experiments, corresponding to the experimental snapshots \textbf{(c,d)}. Since the ice nucleation process is stochastic, it can occur at the solid-liquid interface (a) and near the solid-liquid-air triple point (b). (c) When the ice nucleation occurs at the solid-liquid interface ($c_0=0.15$, $V_0=500~\upmu$L, $T_\text{\text{s}}=-20~^\circ$C), an expanding ice disk can be observed from the top-down view. Subsequently, the migrating ice particles are observed, moving away from the center of the ice disk. (d) When the ice nucleation occurs near the solid-liquid-air triple point ($c_0=0.20$, $V_0=500~\upmu$L, $T_\text{\text{s}}=-20~^\circ$C), the ice disk becomes much smaller than that of the former case. Subsequently, the migrating ice particles are observed, moving towards the other side of the droplet.}
\end{figure*}

The observed interfacial freezing dynamics of the supercooled ethanol-water droplet is quite different from that of the recalescent freezing dynamics of the supercooled water droplet. Given the stochastic nature of the initial ice nucleation process \citep{gurganus2013high}, we capture the freezing dynamics of the large supercooled binary droplet via a top-down view, as shown in Fig.~\ref{fig4} and Movie S2. In order to eliminate the curvature effect of the droplet-air interface on the interfacial freezing dynamics, we perform experiments with large droplets ($V_0=500~\upmu$L), whose initial deposition diameter ($D_0 \approx 2$ cm) is much larger than the capillary length ($l_\text{c}=\sqrt{\gamma/\rho g}\approx2.7$ mm). As a result, gravity flattens the large droplet into a liquid puddle, which has a large aspect ratio of $D_0/h_0 \approx 10$, where $h_0=4V_0/\pi D_0^2\approx 1.6$ mm is the average droplet thickness.

In experiments, according to the initial ice nucleation position, the freezing processes of the supercooled ethanol-water droplet can be categorized into two types: (1) ice nucleation occurs at the solid-liquid interface, as sketched in Fig.~\ref{fig4}(a); (2) ice nucleation occurs near the solid-liquid-air triple point, as sketched in Fig.~\ref{fig4}(b). When the initial ice nucleation position is far from the liquid-air interface ($V_0=500~\upmu$L, $c_0=0.15$), the growth of the circular ice disk at the bottom interface of the droplet is observed, as shown in Fig.~\ref{fig4}(c), which is similar to previous studies on the ice crystal growing on a undercooled substrate \citep{grivet2022contact}. Subsequently, some ice particles form at the droplet-air interface within the affected region of the ice disk, as indicated by the red arrows in Fig.~\ref{fig4}(c). Then, the ice particles migrate and grow at the droplet-air interface, moving away from the center of the ice disk, as indicated by the yellow arrows in Fig.~\ref{fig4}(c). Finally, the ice particles fully cover the droplet-air interface. When the initial ice nucleation position is near the liquid-air interface or at the liquid-air interface ($V_0=500~\upmu$L, $c_0=0.20$), the ice disk at the solid-liquid interface is much smaller than that in the former case, as shown in Fig.~\ref{fig4}(d). Similarly, massive ice particles are observed forming, migrating and growing at the droplet-air interface, moving uni-directionally towards the other side of the droplet, and ultimately covering the entire interface of the supercooled ethanol-water droplet. Varying the initial ethanol concentration $c_0$ (e.g., 0.05, 0.10) yields a similar overall process of ice particle migration and growth at the droplet surface, but with a concentration-dependent velocity to be investigated in the following sections.

%This fascinating interfacial freezing dynamics is still observed when the volume and concentration of the droplet are increased ($V_0=500~\upmu$L, $c_0=0.20$, Fig.~\ref{fig1}c). From the top-down view (Movie S2), the trajectory and the size of ice particles at the interface of supercooled ethanol-water droplets can be tracked much better. Distinct from the dense and continuous ice patterns formed after the recalescence process of supercooled pure water droplets \citep{zhu2022freezing} and supercooled saline droplets \citep{singha2018influence}, hundreds of small dispersed ice particles nucleate progressively, grow slowly and migrate unidirectionally at the droplet-air interface, and ultimately cover the entire interface of the supercooled ethanol-water droplet.

\section{Interfacial temperature evolution of the supercooled binary droplets} 
\label{sec:4}

\begin{figure*}
\centerline{\includegraphics[width=1.0\textwidth]{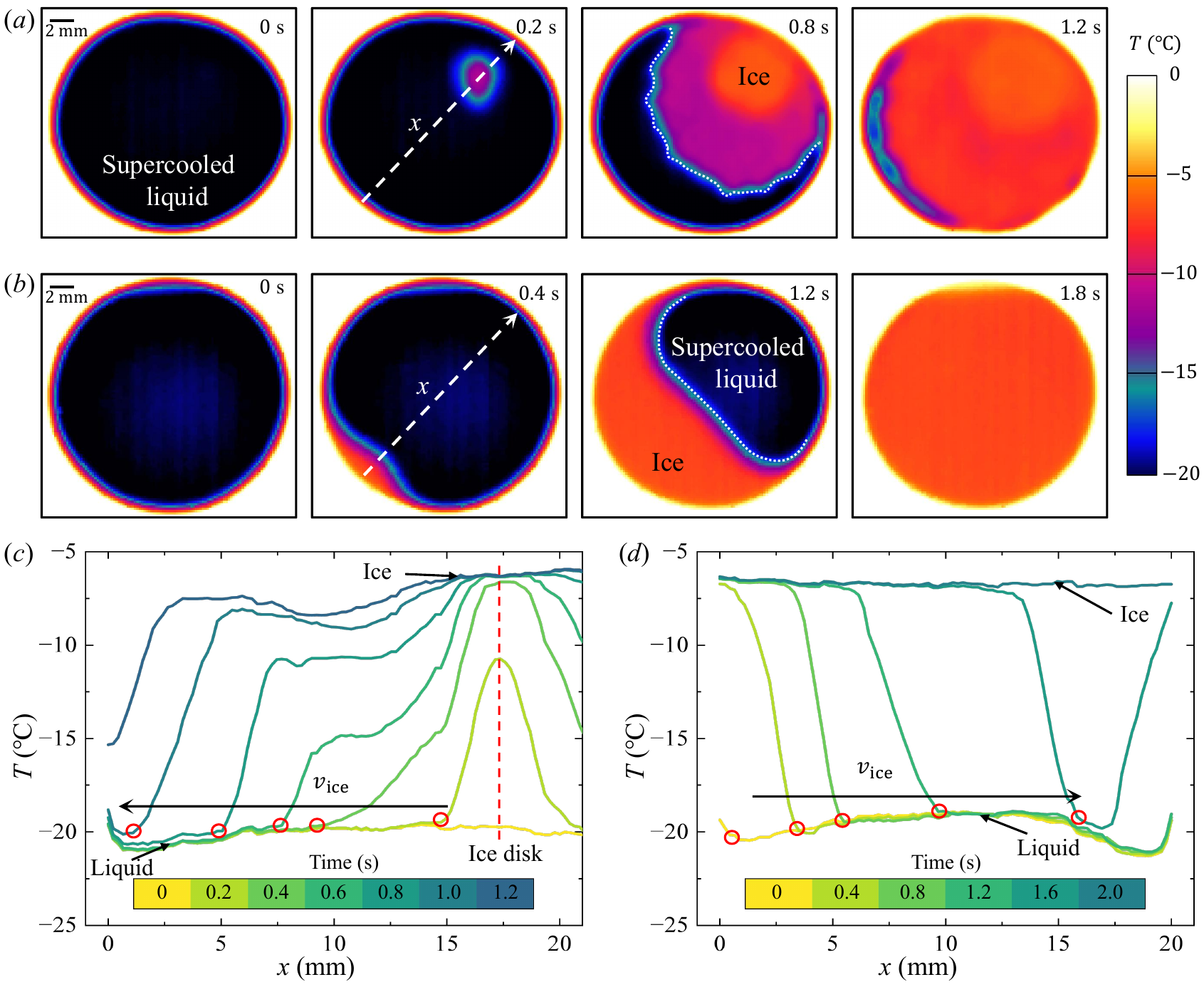}}%
\caption{\label{fig5} Thermal imaging for the freezing supercooled ethanol-water droplets. \textbf{(a)} Top-down view for the case of ice nucleates at the solid-liquid interface ($c_0=0.15$, $V_0=500~\upmu$L, $T_\text{\text{s}}=-20~^\circ$C). \textbf{(b)} Top-down view for the case of ice nucleates near the solid-liquid-air triple point ($c_0=0.20$, $V_0=500~\upmu$L, $T_\text{\text{s}}=-20~^\circ$C). \textbf{(c,d)} Temperature distribution along the white dash line in (a,b), respectively. The temperature of the supercooled liquid $T_\text{d}$ is almost equal to the substrate temperature $T_\text{\text{s}}=-20~^\circ$C. The white dotted lines in (a,b) and the red circles in (c,d) for the ice-water interface.}
\end{figure*}

To understand the mechanism of the interfacial freezing dynamics, we first use the thermal camera. We investigate the interfacial temperature evolution of the supercooled ethanol-water droplets from the top-down view, as shown in Fig.~\ref{fig5}. The initial temperature of droplets ($T_{liq}$) approximately equals the substrate temperature $T_\text{s}$, $T_{liq}\approx T_\text{s}=-20^\circ$C, which is far below the freezing point ($T_\text{m}$), confirming the supercooled state of droplets. When the initial ice nucleation position is far from the liquid-air interface ($V_0=500~\upmu$L, $c_0=0.15$), the growth of the circular ice disk at the bottom interface can also be inferred as well from the thermal imaging snapshot, as indicated by the circular warm region within the droplet in Fig.~\ref{fig5}(a) ($t=0.2$s). Subsequently, when ice particles migrate at the droplet-air interface, due to the released latent heat during the growth of the ice particles, the local temperature near the ice particles increases rapidly to the freezing point ($T_{\text{ice}}=T_{\text{m}}$), as shown in Fig.~\ref{fig5}(a) ($t=0.8$s). Finally, when the ice particles fully cover the droplet-air interface, the interfacial temperature of the droplet approximately equals the freezing point $T_\text{m}$. When the initial ice nucleation position is near the liquid-air interface or at the liquid-air interface ($V_0=500~\upmu$L, $c_0=0.20$), the interfacial temperature evolution is similar to that in the former case.

\begin{figure*}
\centerline{\includegraphics[width=1.0\textwidth]{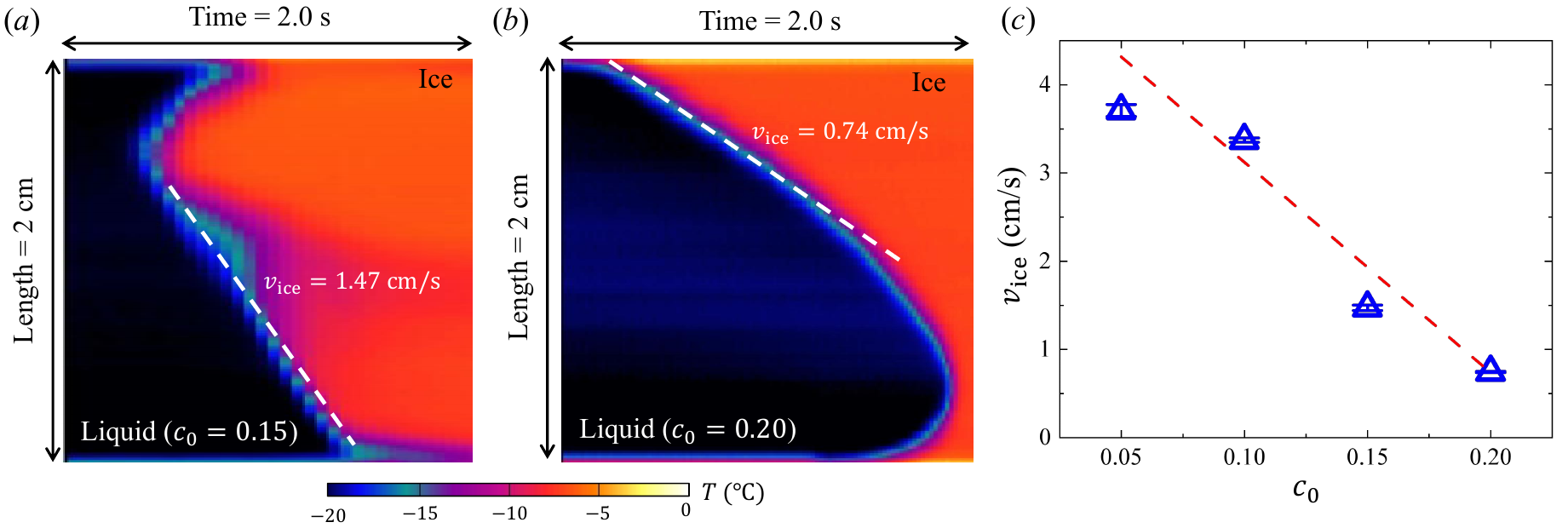}}%
\caption{\label{fig6} Spatial-temporal diagrams of thermal imaging snapshots. \textbf{(a,b)} Spatial-temporal diagrams corresponding to the thermal snapshots in Fig.~\ref{fig5}(a,b), respectively. From the spatial-temporal diagrams, the typical migrating velocity of ice particles $v_\text{\text{ice}}$ can be determined, as indicated by the white dash line. \textbf{(c)} The migrating velocity of ice particles $v_\text{\text{ice}}$ at different ethanol concentrations $c_0$. The blue triangles represent the experimental data, and the red dash line represents the linear fitting result. For the given substrate temperature $T_\text{\text{s}}=-20~^\circ$C, the migrating velocity of ice particles $v_\text{\text{ice}}$ decreases with the ethanol concentration $c_0$. The error bars in (c) for the standard deviations obtained from the linear fits.}
\end{figure*}

To obtain the typical velocity of the interfacial freezing dynamics from the thermal imaging snapshots, we extract the temperature distribution along the centerline of the droplet (Fig.~\ref{fig5}(c,d)), indicated by the white dashed line in Fig.~\ref{fig5}(a,b). Analyzing the transition points of the temperature profiles, indicated by the red circles in Fig.~\ref{fig5}(c,d), yields the typical migrating velocity of ice particles $v_{\text{ice}}$. Furthermore, the temperature profiles in Fig.~\ref{fig5}(c,d) are replotted into spatio-temporal phase diagrams, as shown in \ref{fig6}(a,b). The horizontal axis represents time, and the vertical axis represents space. The slope in the spatiotemporal diagrams, indicated by the white dashed line, determines the typical migrating velocity of ice particles $v_{\text{ice}}$. For the given substrate temperature $T_\text{\text{s}}=-20~^\circ$C, the migrating velocity of ice particles $v_\text{\text{ice}}$ is found to decrease with the ethanol concentration $c_0$.

\section{Marangoni-driven dynamics of ice particles}
\label{sec:5}

In order to fully understand the concentration-related mechanism of the interfacial freezing dynamics, we track the migration and growth of individual ice particles using the open-source software, ImageJ \citep{schneider2012nih}, as sketched in Fig.~\ref{fig7}(a) (details for image processing procedure can be found in Appendix A). The displacement of ice particles $\xi$ as a function of time, at different ethanol concentrations $c_0$, is depicted in Figure~\ref{fig7}(b). From this plot, it is found that the displacement of ice particles $\xi$ increases almost linearly with time. In other words, the ice particles migrate at the droplet-air interface with a constant velocity $v_{\text{ice}}$. By fitting the experimental results with the linear functions indicated by the red solid line in Fig.~\ref{fig7}(b), the migrating velocity of the ice particles $v_{\text{ice}}$ can be determined from the slope. Furthermore, it is found that the migrating velocity $v_{\text{ice}}$ decreases with the ethanol concentration $c_0$, which is consistent with the results obtained from the thermal imaging snapshots in Fig.~\ref{fig6}. Similarly, as shown in Fig.~\ref{fig7}(c), the radius of the ice particles $R_{\text{ice}}$ is also found to increase almost linearly with time. By fitting the experimental results with the linear functions indicated by the red solid line in Fig.~\ref{fig7}(c), the growth rate of the ice particles $\dot R_{\text{ice}}$ can be determined from the slope, and it is also found that the growth rate $\dot R_{\text{ice}}$ decreases with the ethanol concentration $c_0$.

\begin{figure*}
\centerline{\includegraphics[width=1.0\textwidth]{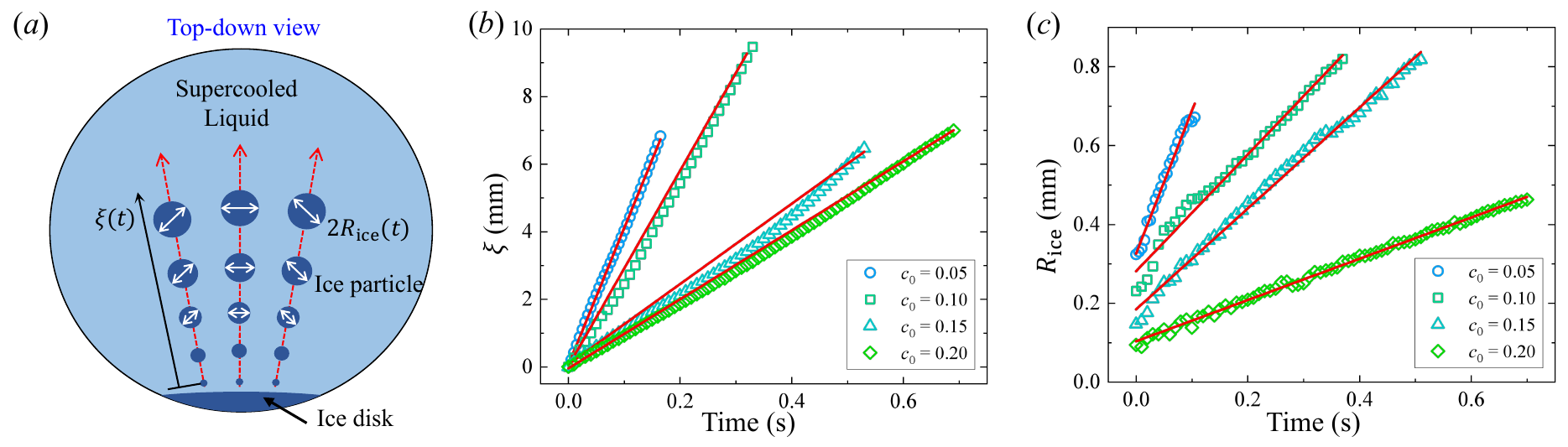}}%
\caption{\label{fig7} Dynamics of migration and growth of the dispersed ice particles at the droplet/air interface. \textbf{(a)} Top-down view sketch for the migration and growth of the dispersed ice particles at the droplet/air interface. \textbf{(b)} The displacement of ice particles $\xi$ and \textbf{(c)} the equivalent radius of ice particles $R_\text{\text{ice}}$ as a function of time, at different ethanol concentrations $c_0$. The dots for the experimental data, and the red solid lines for the linear fitting results.}
\end{figure*}

\subsection{Marangoni-driven migration of ice particles}

\begin{figure*}
\centerline{\includegraphics[width=1.0\textwidth]{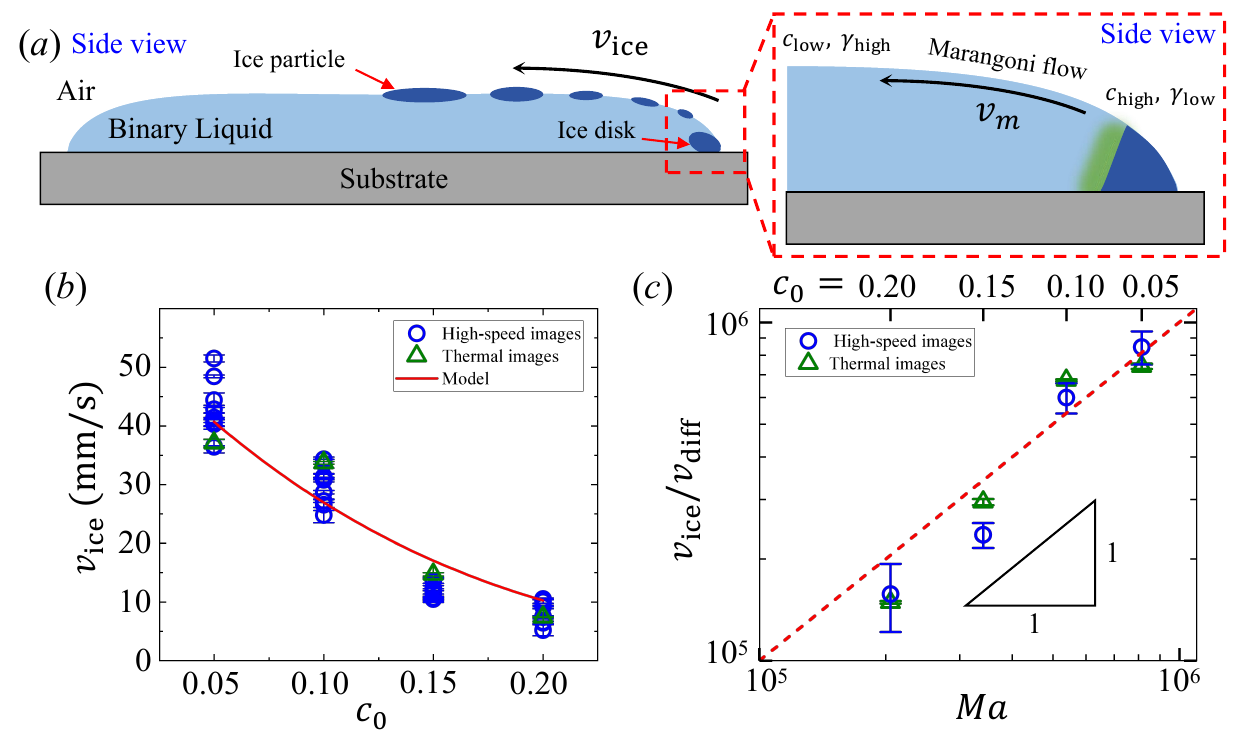}}%
\caption{\label{fig8} Migrating velocity of the ice particles. \textbf{(a)} The side view for the sketch of ice particle migration, and the zoom-in figure is the side view for the sketch of solutal Marangoni flow near the solid-liquid-air triple point. During the formation of the ice disk, the ethanol is rejected from the ice, leading to the local enrichment of ethanol concentration $c_\text{high}$ and the decrease of surface tension $\gamma_\text{low}$. Therefore, the surface tension gradient triggers the solutal Marangoni flow $v_\text{\text{m}}$ away from the triple point, which drives the motion of ice particles. \textbf{(b)} The migrating velocity of ice particles $v_\text{\text{ice}}$ as a function of the ethanol concentration $c_0$. \textbf{(c)} The dimensionless migrating velocity of ice particles $v_\text{\text{ice}}/v_\text{\text{diff}}$ as a function of Marangoni number $Ma$. The dots for the experimental data, the red solid line for the model prediction of $v_{\text{ice}}$, and the red dash line for the scaling relation $v_\text{\text{ice}}/v_\text{\text{diff}} \sim Ma$. The error bars in (b) for the standard deviations obtained from the linear fits, and the error bars of blue circles in (c) for the standard deviation calculated from at least ten data points.}
\end{figure*}

Considering the small Stokes number of the migrating ice particles $Stk=\rho v_{\text{ice}} R_{\text{ice}}/18 \mu < 0.1$, we assume that the migration velocity of the ice particles $v_{\text{ice}}$ is determined by the solutal Marangoni flow at the droplet-air interface, as sketched in Fig.~\ref{fig8}(a). During solidification of supercooled ethanol-water droplets, the local concentration of ethanol $c_\text{high}$ in the vicinity of the ice front increases, reaching the saturation concentration $c_\text{sat}$, at which $T_\text{m}(c_\text{sat})=T_\text{s}$. Consequently, the surface tension $\gamma _\text{low}$ near the ice front is smaller than that away from the ice front $\gamma_\text{high}$, which triggers the solutal Marangoni flow across the droplet and drives ice particles to move \citep{lohse2020physicochemical}. Then, the migration velocity $v_{\text{ice}}$ can be estimated by balancing the Marangoni stress $\Delta \gamma /D_0$ with the viscous stress $\mu v_{\text{m}}/l$, 
\begin{equation}
v_{\text{ice}}=v_{\text{m}}=\frac{\Delta \gamma l}{\mu D_0},
\label{vice}
\end{equation}
where $\mu$ is the viscosity of the binary mixture at the saturated concentration $c_\text{sat}=0.37$, $D_0=2$ cm is the droplet diameter, $l=h_0/2$ is the typical length scale, taken as half the average droplet thickness $h_0=V_0/\pi R_0^2\approx 1.6$ mm to account for the compensating flow near the substrate \citep{gelderblom2022evaporation}. $\Delta \gamma=\gamma _{0}-\gamma _\text{sat}$ is the typical surface tension difference, the surface tension far from the ice front $\gamma _0$ remains the initial state, which is determined by the initial concentration $c_0$, and the surface tension near the ice front $\gamma _\text{sat}$ is determined by the saturated state, determined by the saturated concentration $c_\text{sat}$, as listed in Tab.~\ref{tab:my_label}. It should be noted that the ice formed in binary liquid system is considered to be porous \citep{wei2025hybrid}, which is similar to the sea ice \citep{du2023sea,du2025sea}. While the ice phase consists solely of water, the porous ice is filled with the concentrated ethanol-water mixture. Therefore, we assume a saturated concentration near the ice front. 

After incorporating the concentration-dependent properties of the ethanol-water mixture, the estimated ice particle migration velocity $v_{\text{ice}}$ (Eq.~\ref{vice}) agrees well with the experimental observations, as shown in Fig.~\ref{fig8}(b). The minor discrepancies can likely be attributed to several factors: (1) potential contamination by impurities or surfactants, as reported by \citet{marin2016surfactant,marin2019solutal,diddens2017detailed}; (2) the finite thickness of the ice particles, which reduces the liquid film thickness above the substrate and increases viscous dissipation; (3) the use of liquid properties (surface tension and density) measured at room temperature, while the experiments were conducted under an undercooling of $T=-20~^\circ$C; (4) possible collective effects arising from interactions between multiple ice particles; (5) the limited applicability of the model's simplifying assumptions as the ice particles grow larger and migrate away from the source toward the droplet edge.

Qualitatively, for a given substrate temperature, the migration velocity $v_{\text{ice}}$ is found to decrease with concentration $c_0$ as shown in Fig.~\ref{fig8}(b), which arises from the concentration-related surface tension difference $\Delta \gamma (c_0)$. Considering $\Delta \gamma (c_0)=\gamma _{0}(c_0)-\gamma _\text{sat}$, the initial surface tension $\gamma _0$ decreases with concentration $c_0$, while the saturated surface tension $\gamma _\text{sat}$ is almost constant for the given substrate temperature $T_\text{s}$. Thus, both $\Delta \gamma$ and $v_{\text{ice}}$ decrease with concentration $c_0$.

Additionally, it should be noted that, although the temperature gradient in Fig.~\ref{fig5} might trigger the thermal Marangoni flows to drive the ice particle to move, the thermal Marangoni effect is much weaker than the solutal Marangoni effect in our experiments. The typical surface tension difference induced by temperature $\Delta \gamma _T$ can be estimated by $\Delta \gamma _T \sim |\text{d}\gamma/\text{d}T|\cdot\Delta T < 2$ mN/m, where $|\text{d}\gamma/\text{d}T| \approx 2 \times 10^{-4}$ N/(m$\cdot$K), and the temperature difference $\Delta T=T_\text{m}-T_\text{s} < 20$ K. While the typical surface tension difference induced by the ethanol concentration $\Delta \gamma _\text{c}$ should be on the order of 10 mN/m, as listed in Tab.~\ref{tab:my_label}. As a result, the thermal Marangoni effect on interfacial freezing dynamics could be ignored in the present work.

Furthermore, the migrating velocity of the ice particles $v_{\text{ice}}$ is non-dimensioned by the typical velocity of solute diffusion $v_{\text{diff}}=\mathscr{D}/D_0$, where $\mathscr{D}=1\times 10^{-9}$m$^2$/s is the mass diffusivity of ethanol in water and $D_0$ is the diameter of the droplet. Then, the dimensionless migrating velocity $v_{\text{ice}}/v_{\text{diff}}$ can be described by the solutal Marangoni number $Ma$,
\begin{equation}
\frac{v_{\text{ice}}}{v_{\text{diff}}} = \frac{\Delta \gamma h}{2\mu \mathscr{D}} = Ma.
\label{Marangoni}
\end{equation}

This scaling relation (Eq.~\ref{Marangoni}) agrees well with the experimental data, as shown in Fig.~\ref{fig8}c. Indeed, the dimensionless migrating velocity $v_{\text{ice}}/v_{\text{diff}}$ also reflects the effective Peclet number $Pe$ at the droplet-air interface, $Pe=D_0 v_{\text{ice}}/\mathscr{D}=v_{\text{ice}}/v_{\text{diff}}$. Hence, the large dimensionless migrating velocity $v_{\text{ice}}/v_{\text{diff}} \gg 1$ in Fig.~\ref{fig8}c suggests the effective Peclet number $Pe \gg 1$, and convection is dominant at the droplet-air interface.

\subsection{Marangoni-driven growth of ice particles}
\begin{figure*}
\centerline{\includegraphics[width=1.0\textwidth]{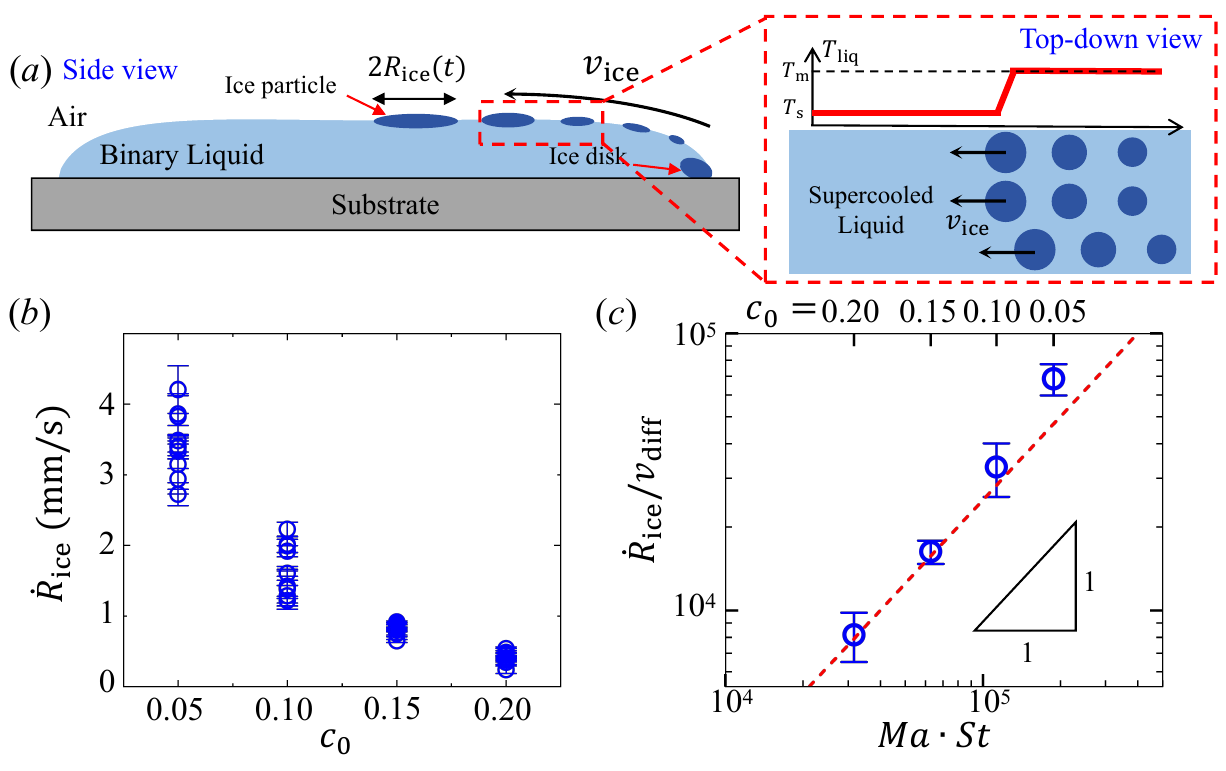}}%
\caption{\label{fig9} Growth rate of the ice particles. \textbf{(a)} The side view for the sketch of ice particle growth, and the zoom-in figure is the top-down view for the sketch of the temperature distribution around the ice particle. According to the thermal snapshots, the liquid ahead of migrating ice particles remains supercooled. Therefore, the heat transfer within the supercooled liquid predominantly consume the released latent heat during ice particle growth. \textbf{(b)} The growth rate of ice particles $\dot{R}_\text{\text{ice}}$ as a function of the ethanol concentration $c_0$. \textbf{(c)} The dimensionless migrating velocity of ice particles $\dot{R}_\text{\text{ice}}/v_\text{\text{diff}}$ as a function of Marangoni number $Ma$ and Stefan number $St$. The dots for the experimental data, the red solid lines for the model predictions, and the red dash line for the scaling relation $\dot{R}_\text{\text{ice}}/v_\text{\text{diff}} \sim Ma \cdot St$. The error bars in (b) for the standard deviations obtained from the linear fits, and the error bars in (c) for the standard deviation calculated from at least ten data points.}
\end{figure*}

Similarly, we turn to analyzing the physical mechanism of ice particle growth. As sketched in Fig.~\ref{fig9}(a), ice particles migrate uni-directionally with a constant velocity $v_{\text{ice}}$ at the droplet-air interface. For the leading ice particles (as sketched in the red dashed box), the interfacial temperature behind the ice particles approximately equals the freezing point $T_\text{m}$, while the interfacial temperature ahead of the ice particles remains supercooled $T_{liq}=T_\text{s}<T_\text{m}$, as revealed by the thermal snapshots in Fig.~\ref{fig5}. The growth dynamics of ice particles can therefore be modeled with the classical two-phase Stefan problem \citep{ahmadi2019soap,huerre2021solidification}, in which the latent heat is gradually dissipated into the heat sink (the supercooled liquid) to keep the system at or just below the freezing point. By balancing the thermal fluxes at the ice-water interface, the growth rate of the ice particles $\dot R_{\text{ice}}$ can be expressed as follows,
\begin{equation}
\rho _\text{l} L_\text{f} \dot R_{\text{ice}} = k_\text{i} \nabla T_\text{i}-k_\text{l} \nabla T_\text{l},
\label{latent}
\end{equation}
where $\rho _\text{l}$ and $L_\text{f}$ are the density of liquids and the latent heat of solidification, $k_\text{i}, k_\text{l}$ are the thermal conductivities of the ice phase and the liquid phase, $T_\text{i}, T_\text{l}$ are the temperatures within the ice phase and the liquid phase.

Since the ice particles migrate along the droplet–air interface, the released latent heat is primarily dissipated into the supercooled liquid (see the last term in Eq.~\ref{latent}). Consequently, we need to account for the convective heat transfer induced by the particle migration velocity $v_\text{ice}$. Considering the quasi-static advection-diffusion equation,
\begin{equation}
v_\text{ice} \nabla T_\text{l}=\alpha_\text{l} \nabla ^2 T_\text{l},
\label{convection}
\end{equation}
where $\alpha _\text{l}=k _\text{l} /\rho _\text{l} c_p$ is the thermal diffusivity coefficient of the liquid, $c_p$ is the specific heat capacity of liquid. 

Assuming a typical length scale $l^*$, Eq.~\ref{latent} simplifies to $\rho _\text{l} L_\text{f} \dot R_{\text{ice}}=k_\text{l}  \Delta T/l^*$, and Eq.~\ref{convection} reduces to $v_\text{ice}\Delta T/l^*=\alpha _\text{l}\Delta T/{l^*}^2$. Then, by combining Eq.~\ref{latent} and Eq.~\ref{convection} to eliminate $l^*$, the growth rate of the ice particles $\dot R_{\text{ice}}$ can be expressed as follows,
\begin{equation}
\dot R_{\text{ice}}=\frac{ v_{\text{ice}} k_\text{l} \Delta T}{\alpha _\text{l} \rho _\text{l} L_\text{f}}.
\label{growth}
\end{equation}

Qualitatively, the growth rate of the ice particles $\dot R_{\text{ice}}$ is also found to decrease with concentration $c_0$, as shown in Fig.~\ref{fig9}(b). The reason is that for a given substrate temperature $T_\text{s}$, with increasing concentration $c_0$, the freezing point of the binary droplets $T_\text{m}$ decreases, and thus the supercooling temperature decreases $\Delta T(c_0)=T_\text{m}(c_0)-T_\text{s}$. Thus, both $v_{\text{ice}}$ and $\Delta T$ decrease with $c_0$, and consequently $\dot R_{\text{ice}}$ decreases more drastically with concentration $c_0$ (Eq.~\ref{growth}), compared with $v_{\text{ice}}$ (Fig.~\ref{fig8}(b)).

The growth rate of the ice particles $\dot R_{\text{ice}}$ is also non-dimensioned by the typical velocity of solute diffusion $v_{\text{diff}}=\mathscr{D}/D_0$. Then, the dimensionless growth rate of ice particles $\dot R_{\text{ice}}/v_{\text{diff}}$ can be described by the solutal Marangoni number $Ma$ and the Stefan number $St$,
\begin{equation}
\frac{\dot R_{\text{ice}}}{v_{\text{diff}}}=\frac{v_{\text{ice}} D_0}{\mathscr{D}} \cdot \frac{c_p \Delta T}{L_\text{f}} \sim Ma \cdot St,
\label{grow}
\end{equation}
where $St=c_p \Delta T/L_\text{f}$ is the Stefan number \citep{zeng2022influence}. This scaling relation (Eq.~\ref{grow}) again agrees well with the experimental data, as shown in Fig.~\ref{fig9}(c). Moreover, in the present experiments, evaporative cooling of the ethanol-water mixture has little effect on the reported interfacial freezing dynamics (see detailed analysis in Appendix B). By contrast, when a highly volatile liquid (e.g., ether) is mixed with another whose melting point is near room temperature (e.g., hexadecane), selective evaporation can release enough latent heat to freeze the droplet, as demonstrated by \citet{kant2020pattern} and \citet{zeng2023evaporation}.

\subsection{Final wrapping state of the binary droplet}

\begin{figure*}
\centerline{\includegraphics[width=1.0\textwidth]{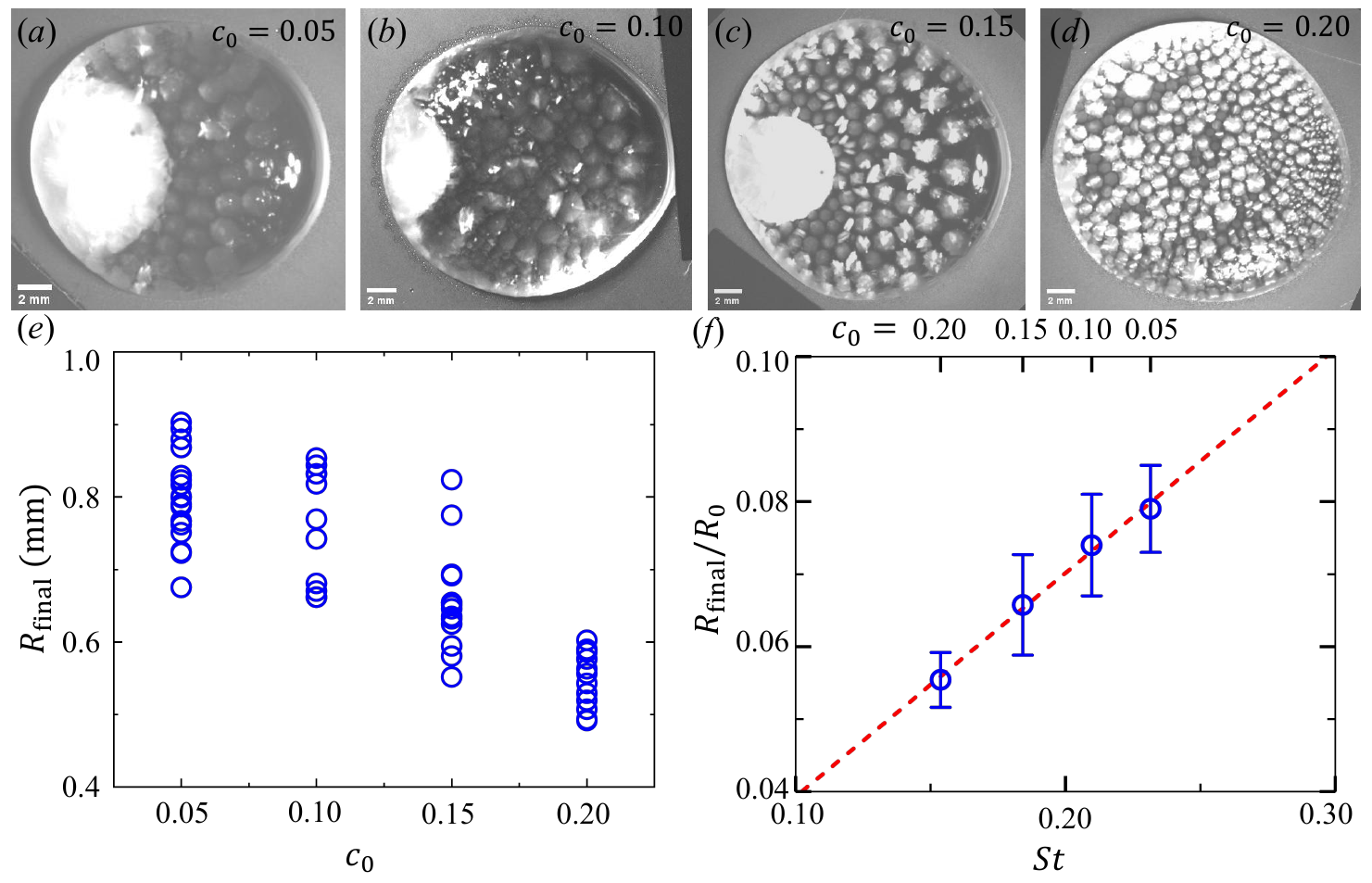}}%
\caption{\label{fig10} Final wrapping state of the frozen supercooled ethanol-water droplet. \textbf{(a-d)} Top-down view for the final wrapping state of the frozen supercooled ethanol-water droplets at different ethanol concentrations $c_0$. \textbf{(e)} Equivalent radius of ice particles $R_\text{final}$ at the final wrapping state as a function of ethanol concentrations $c_0$. \textbf{(f)} Dimensionless radius of ice particles $R_\text{final}/R_0$ at the final wrapping state as a function of Stefan number $St$. The dots for the experimental data, the red solid lines for the model predictions, and the red dash line for the linear relation between $R_\text{final}/R_0$ and $St$. The error bar for the standard deviation of at least fifteen data points.}
\end{figure*}

Based on our understanding of ice particle dynamics, we now focus on the final wrapping state of freezing supercooled ethanol-water droplets, when the droplet is fully covered by dispersed ice particles, as shown in Fig.~\ref{fig10}(a-d). With increasing concentration $c_0$, the final radius $R_\text{final}$ of the ice particles is found to decrease, while the thickness $h_{\text{ice}}$ of the ice particles increases, which can be inferred from the transparency of ice particles. The dependence of final wrapping state on concentration $c_0$ can be explained by considering the characteristic timescale $\tau_\text{c}$ for ice particles traveling across the droplet, $\tau_\text{c} \sim R_0/v_{\text{ice}}$, where $R_0 \approx 1$ cm is the initial radius of droplet in experiments for all $c_0$. Then, the equivalent radius of ice particles $R_\text{final}$ at the final wrapping state can be expressed as follows,
\begin{equation}
R_\text{final}=\tau_\text{c} \cdot \dot R_{\text{ice}} \sim \frac{c_p\Delta T R_0}{L_\text{f}}.
\label{Rfinal}
\end{equation}

The predictions of the final radius of ice particles $R_\text{final}$ well describe the trend of the experimental data, as shown in Fig.~\ref{fig10}(e). From Eq.~\ref{Rfinal}, the final radius of the ice particles $R_\text{final}$ is found to be solely determined by the supercooling temperature $\Delta T$. Therefore, the freezing point depression effect of the ethanol concentration $c_0$ results in the final radius of the ice particles $R_\text{final}$ decreasing with concentration $c_0$. Then, the final radius $R_\text{final}$ is further non-dimensioned by the radius of the droplet $R_0$, and the dimensionless radius of ice particles at the final state $R_\text{final}/R_0$ can be expressed as follows,
\begin{equation}
\frac{R_\text{final}}{R_0} = \frac{\tau_\text{c} \cdot \dot R_{\text{ice}}}{R_0} \sim St.
\label{final}
\end{equation}

This scaling relation (Eq.~\ref{final}) is further validated by the experimental data, as shown in Fig.~\ref{fig10}(f). Besides, as the concentration $c_0$ decreases, the characteristic timescale $\tau_\text{c}=R_0/v_{\text{ice}}$ also decreases. If we assume that the ice growth along the thickness direction obeys the diffusive growth dynamics $h_{\text{ice}} \sim \sqrt{\tau_\text{c}}$ \citep{ghabache2016frozen}, then the ice particles would be thinner and consequently more transparent as the concentration $c_0$ decreases.

\section{Conclusion and discussions}
\label{sec:6}

In conclusion, we present a compelling study on the interfacial freezing dynamics of supercooled ethanol-water droplets, revealing the critical role of the internal solutal Marangoni flow, triggered by the localized enriched concentration of ethanol near the ice front. By incorporating the fluid's property variations, our model quantitatively predicts the ice particle's migration velocity via the Marangoni flow and its growth rate via a balance of latent heat and solutal-Marangoni-enhanced thermal dissipation.

Our findings may advance the fundamental understanding of complex physicochemical hydrodynamics in the context of multicomponent liquids undergoing phase transitions and may further pave the way for droplet-based technological innovations in manufacturing and material science \citep{lohse2020physicochemical,gao2025characterization,wang2025physics}. More efforts are needed to understand the morphological evolution and instability of the solidification front within multicomponent liquid systems, spanning from the molecular scale \citep{wahl2020ice,facq2013ice} to the nanometer scale \citep{ahmadi2019soap,grivet2022contact} and further to the macroscopic scale \citep{anderson2020convective,du2025sea}, especially when the solidification process is coupled with other physical processes, such as flash evaporation \citep{kant2020pattern,zeng2023evaporation}.

%\backsection[Supplementary data]{\label{SupMat}Supplementary material and movies are available at \\https://doi.org/10.1017/jfm.2019...}

\backsection[Acknowledgements]{We thank the anonymous reviewers for their insightful comments, which have greatly helped improve the clarity of this work.}

\backsection[Funding]{This work is supported by NSFC Excellence Research Group Program for ‘Multiscale Problems in Nonlinear Mechanics’ (No. 12588201), the Natural Science Foundation of China under Grant No. 12402321, the National Key R\&D Program of China under Grant No. 2021YFA0716201, the New Cornerstone Science Foundation through the New Cornerstone Investigator Program and the XPLORER PRIZE, the Postdoctoral Fellowship Program of the China Postdoctoral Science Foundation under Grant Nos. 2025T180524, GZB20240366 and 2024M751637, and Shuimu Tsinghua Scholar Program under Grant No. 2023SM038.}

\backsection[Declaration of interests]{ The authors report no conflict of interest.}

\backsection[Author ORCID]{\protect\\
Feng Wang, https://orcid.org/0000-0003-0025-3895\\
Hao Zeng, https://orcid.org/0000-0003-3190-5892\\
Yihong Du, https://orcid.org/0009-0004-2162-7037\\
Chao Sun, https://orcid.org/0000-0002-0930-6343}

\section*{Appendix A: Imaging processing procedure}

\begin{figure*}
\centerline{\includegraphics[width=1.0\textwidth]{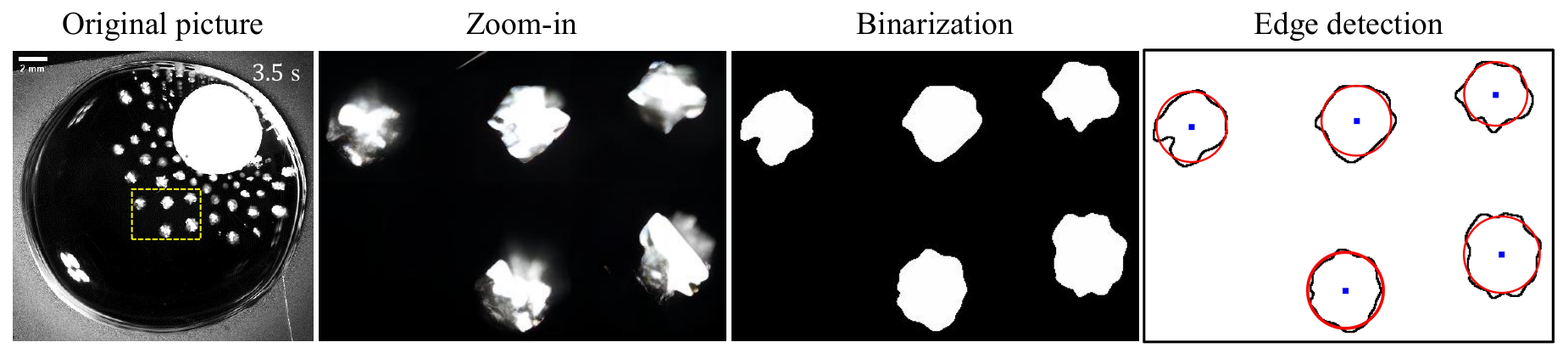}}%
\caption{\label{fig11}  Image-processing procedure for analyzing the displacement $\xi$ and the radius $R_\text{\text{ice}}$ of the ice particles. For a typical snapshot ($t=3.5$s in Fig.~\ref{fig4}(c)), the original picture first undergoes binarization and denoising. Then, using edge-detection methods, the outline of the ice particle is obtained. Subsequently, we can calculate the area $S$ and the mass center of the ice particle $(x,y)$. Then, we calculate the equivalent radius $R_\text{\text{ice}}=\sqrt{S/\pi}$, under the assumption of the circular shape of the ice particle. We can also calculate the displacement $\xi=\sqrt{(x-x_0)^2+(y-y_0)^2}$ of the ice particle, where $(x_0,y_0)$ is the initial position of the ice particle.}
\end{figure*}

For a typical snapshot ($t=3.5$s in Fig.~\ref{fig4}(c)), the original picture (the yellow dashed box) first undergoes binarization and denoising, as shown in Fig.~\ref{fig11}. The white region represents the ice particle. Then, we use edge-detection methods to obtain the outline of the ice particle, as indicated by the black solid line in the last image of Fig.~\ref{fig11}. Subsequently, we can calculate the area $S$ and the mass center of the ice particle $(x,y)$, as indicated by the blue dots in the last image of Fig.~\ref{fig11}. Then, we calculate the equivalent radius $R_\text{\text{ice}}=\sqrt{S/\pi}$, under the assumption of the circular shape of the ice particle, as indicated by the red circles. Additionally, we can also calculate the displacement $\xi=\sqrt{(x-x_0)^2+(y-y_0)^2}$ of the ice particle, where $(x_0,y_0)$ is the initial position of the ice particle. Using these image-processing and data-processing procedures, the migration and growth dynamics of a single ice particles can be extracted from the experimental snapshots, as sketched in Fig.~\ref{fig7}(a).

\section*{Appendix B: Discussions of the evaporation-cooling effect}
\begin{figure*}
\centerline{\includegraphics[width=0.6\textwidth]{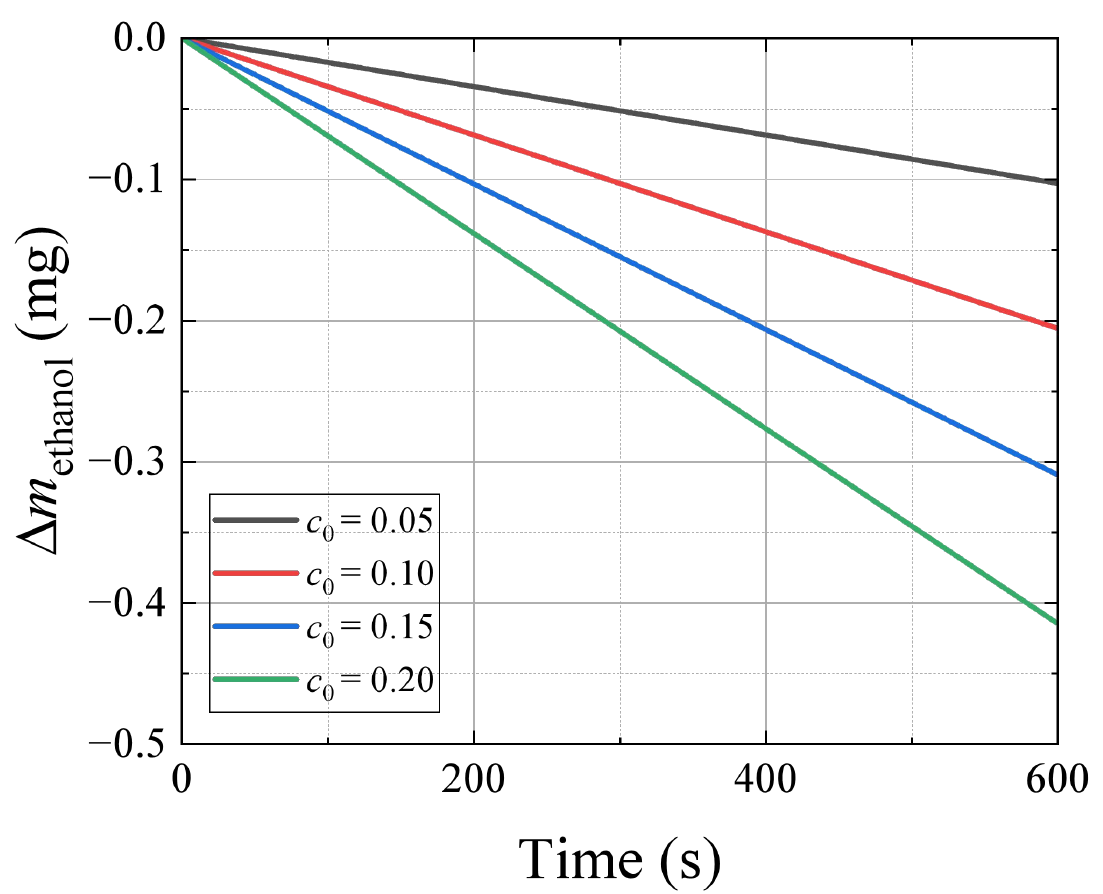}}%
\caption{\label{fig12} Ethanol evaporation rate of ethanol-water liquid films (initial volume $V_0=150~\mu$L, radius $R_0=1$ cm)  with various ethanol fractions.}
\end{figure*}

To assess the influence of droplet evaporation on the observed interfacial freezing dynamics, we compute the evaporation rate of ethanol-water films (initial volume $V_0=500~\mu$L, radius $R_0=1$ cm) with different ethanol fractions using the numerical models reported by \citet{zeng2023evaporation} and \citet{wakata2024evaporation,wakata2025thermal}. Figure~\ref{fig12} shows the ethanol evaporation rate for films with various ethanol fractions ($c_0=0.05, 0.10, 0.15, 0.20$). The typical ethanol evaporation flux can be estimated as $\dot{m}_\text{ethanol} \sim 0.4~\text{mg}/600~\text{s} \sim 10^{-9} ~\text{kg/s}$. Therefore, the heat flux induced by ethanol evaporation can be estimated as follows,
\begin{equation}
Q_\text{evap}=L_e \dot{m}_\text{ethanol} \sim 10^{-6}~\text{kJ/s}, 
\end{equation}
where $L_e=841$ kJ/kg is the latent heat of vaporization of ethanol.

By contrast, the typical growth rate of the ice particles $\dot{R}_\text{ice} \sim 1~\text{mm/s}$, so the released latent heat of ice particle growth can be estimated as follows,
\begin{equation}
Q_\text{ice}=L_f \rho _\text{ice}\pi R_0 \dot{R}_\text{ice} \sim 10^{-1}~\text{kJ/s},
\end{equation}
where $\rho _\text{ice}=917~\text{kg/m}^3$ is the density of ice.

The released latent heat of ice particle growth is much larger than the heat flux induced by ethanol evaporation, $Q_\text{ice} \gg Q_\text{evap}$. Therefore, ethanol evaporation has a negligible effect on the reported interfacial freezing dynamics. 

%\bibliographystyle{jfm}
%\bibliography{jfm}

%Use of the above commands will create a bibliography using the .bib file. Shown below is a bibliography built from individual items.

\bibliographystyle{jfm}

%% End of file `jfm2esam.bib'.

\end{document}